\documentclass[aps,prd,onecolumn,groupedaddress,showpacs,nofootinbib,amssymb]{revtex4-2}
\usepackage[dvips]{graphicx}
\usepackage{amssymb}
\usepackage{amsmath}
\usepackage{graphicx,,color}
\usepackage{amsfonts}
\usepackage{bm}
\usepackage{cancel}
\usepackage{comment}
\usepackage{floatflt}
\usepackage{slashed}
\newcommand\be{\begin{equation}}
\newcommand\ee{\end{equation}}

\allowdisplaybreaks[4]

\begin{document}

\title{Axion-Neutrino Couplings, Late-time Phase Transitions and the Far Infrared Physics}
\author{V.K. Oikonomou,$^{1,2}$}\email{voikonomou@gapps.auth.gr;v.k.oikonomou1979@gmail.com}
\affiliation{$^{1)}$Department of Physics, Aristotle University of
Thessaloniki, Thessaloniki 54124, Greece \\ $^{2)}$L.N. Gumilyov
Eurasian National University - Astana, 010008, Kazakhstan}

%$^{2)}$ Laboratory for Theoretical Cosmology, International Center
%of Gravity and Cosmos, Tomsk State University of Control Systems
%and Radioelectronics  (TUSUR), 634050 Tomsk, Russia

 \tolerance=5000

\begin{abstract}
The far infrared physics is a fascinating topic for theoretical
physics, since the foundation of quantum field theory and
neutrinos seem to be strongly related with the far infrared
physics of our Universe. In this work we shall explore the
possibility of a late-time thermal phase transition caused by the
axion-neutrino interactions. The axion is assumed to be the
misalignment axion which is coupled primordially to a chiral
symmetric neutrino. The chiral symmetry is supposed to be broken
either spontaneously or explicitly, and two distinct
phenomenological models of axion-neutrinos are constructed. The
axion behaves as cold dark matter during all its evolution eras,
however if we assume that the axion and the neutrino fields
interact coherently in a classical way as fields, or as ensembles,
then we consider thermal effects in the axion sector, due to the
values of operators $\phi$ for the axion and $\bar{\nu}\nu$ due to
the neutrinos. The thermal equilibrium between the two has no
effect to the axion effective potential for a wide temperature
range. As we show, contrary to the existing literature, the axion
never becomes destabilized due to the finite temperature effects,
however if axion-Higgs higher order non-renormalizable operators
are present in the Lagrangian, the axion potential is destabilized
in the temperature range $T\sim 0.1\,$MeV down to $T\sim 0.01\,$eV
and a first order phase transition takes place. The initial axion
vacuum decays to the energetically more favorable axion vacuum,
and the latter decays to the Higgs vacuum which is more
preferable. This late-time phase transition might take place in
the redshift range $z\sim 385-37$ and thus it may cause density
fluctuations in the post-recombination era. This might be the
source of large scale matter structure at high redshifts $z\geq
9$. Following the literature, we qualitatively discuss the
implications of such a late-time phase transition at the
astrophysical and cosmological level.
\end{abstract}

%PACS numbers: 04.50.Kd, 95.36.+x, 98.80.-k, 98.80.Cq
\pacs{04.50.Kd, 95.36.+x, 98.80.-k, 98.80.Cq,11.25.-w}

\maketitle

\section{Introduction}

The far infrared  physics is an unexplored scientific area, and is
not accessible in accelerator experiments. However, the infrared
physics might play a crucial role in nature, especially at late
times in the post Cosmic Microwave Background (CMB) era. An
intriguing coincidence motivating such way of thinking is the fact
that the Universe's vacuum energy is of the order
$(0.003\,\mathrm{eV})^4$, while the absolute rest mass of the
neutrino is believed to be $m_{\nu} \sim
\mathcal{O}(10^{-1}-10^{-3})\,$eV, hence it seems that loosely
speaking, the current vacuum energy of the Universe is of the
order $\sim \mathcal{O}(m_{\nu}^4)$. These arguments are rather
theoretical motivations based on infrared physics which fascinate
theoretical physicists with deep knowledge of quantum field theory
foundations. Apart from these arguments, there are by far more
compelling reasons on why should late-time infrared physics be
more complicated in comparison to the standard
$\Lambda$-Cold-Dark-Matter ($\Lambda$CDM) model. The most
important reason is the possible existence of large scale matter
structure at high redshifts, higher than $z=6$, and the tensions
of the $\Lambda$CDM model between late-time matter sources and the
CMB, including the clustering problem of the $\Lambda$CDM. There
exist indications of large scale structure in high redshifts, and
it is usually said that this would put in peril the $\Lambda$CDM
model, usually with some desperate expressions. But the
$\Lambda$CDM model is not the actual description of nature at late
times, it is a general relativistic framework with basic dark
matter which seems to be compatible with the CMB polarization
anisotropies. It is basically a cosmological constant with dark
matter, so no one knows if general relativity is sufficient to
describe late-time physics, for example modified gravity
\cite{reviews1,reviews2,reviews3,reviews4,reviews5} can also mimic
the $\Lambda$CDM and provide a dynamical dark energy component in
the Universe. Regardless of that perspective, large scale matter
structure at high redshifts is rather difficult to be explained
with a standard nearly scale invariant power spectrum of
primordial scalar perturbations and the $\Lambda$CDM.

There exists an appealing possibility for explaining the high
redshift matter structure, already proposed in the literature, the
late-time phase transitions which might have occurred in the
post-CMB era
\cite{Kolb:1992uu,Frieman:1991tu,Fuller:1990mq,Primack:1980ve,Wasserman:1986gi,Hill:1988vm,Press:1989id,Luo:1993tg,Dutta:2009ix,Ramberg:2022irf,Patwardhan:2014iha}.
Indeed, such late-time phase transitions might lead to density
fluctuations in the post-CMB era, which can serve for seeds of
high redshift matter structure
\cite{Wasserman:1986gi,Patwardhan:2014iha}. Such late-time phase
transitions do not disturb the CMB temperature anisotropy $\delta
T/T$, and therefore the CMB last scattering remains smooth. If
these phase transitions occurred earlier than the CMB, we would
see remnants of this transition in the CMB photons. It is
remarkable that late-time phase transitions may lead to structure
formation at redshifts which belong to a forbidden region of the
$\Lambda$CDM model and to mass scales up to $M\sim
10^{18}M_{\odot}$ \cite{Hill:1988vm}, although such massive
objects have never be observed. These massive objects if they
exist, will be a challenge for the $\Lambda$CDM model to explain.

In this line of research, in this paper we shall explore the
possibility that neutrinos interact primordially with the
misalignment axion
\cite{Preskill:1982cy,Abbott:1982af,Dine:1982ah,Marsh:2015xka,Sikivie:2006ni,Raffelt:2006cw,Linde:1991km,Co:2019jts,Co:2020dya,Barman:2021rdr,Marsh:2017yvc,Odintsov:2019mlf,Odintsov:2019evb,maxim,Anastassopoulos:2017ftl,Sikivie:2014lha,Sikivie:2010bq,Sikivie:2009qn,Masaki:2019ggg,Soda:2017sce,Soda:2017dsu,Aoki:2017ehb,Arvanitaki:2019rax,Arvanitaki:2016qwi,Machado:2019xuc,Tenkanen:2019xzn,Huang:2019rmc,Croon:2019iuh,Day:2019bbh,Oikonomou:2022ela,Oikonomou:2022tux,Odintsov:2020iui,Oikonomou:2020qah,Oikonomou:2023bah,DiLuzio:2020wdo,Visinelli:2018utg,Mazde:2022sdx,Lambiase:2022ucu,Ramberg:2019dgi}.
We shall consider two models which both have a primordial chiral
symmetry, which is either broken explicitly or spontaneously. The
former case is aligned with the naturalness argument and the axion
mass occurs due to fermion one-loop corrections. In both cases,
the axion is thermalized via its interaction with the neutrino, in
a classical way, if the axion and the neutrino are considered as
classical particle/field ensembles, in a coherent interaction
between the axion and neutrino fields. The thermalization of the
axion does not occur in a microphysical way, via its interactions
with the neutrino and the corresponding decay rates, its a
coherent effect. One can consider such interaction as a collective
classical interaction of the coherent axion oscillations with the
neutrino thermal bath. If we accept physically this procedure, we
find some remarkable results. Specifically, when specific higher
order operators with axion-Higgs interactions are considered
\cite{Oikonomou:2023bah}, we show that once the neutrino decouples
from the electroweak sector at a temperature $T\sim 1\,$MeV phase
transitions occur in the axion-neutrino sector, in the temperature
range $T\sim 1\,$MeV to $T\sim 0.01\,$eV, with the lower bound
being assumed to be absolute neutrino rest mass. When the
temperature is of the same order as the neutrino mass, the
neutrino decouples from the axion-neutrino thermal equilibrium.
Such thermal phase transitions have been studied in the
literature, however as we point out there is no thermal phase
transition in this system without the Higgs-axion operators. In
fact, it proves that the neutrino sector plays a crucial role in
this late-time thermal phase transition, which proves to be first
order. Accordingly, we discuss the astrophysical and cosmological
implications of such a late-time thermal phase transition in the
neutrino-axion sector.

This paper is organized as follows: In section II we briefly
review the misalignment axion mechanism and we discuss the role of
the axion as a dark matter particle. In section III we develop the
axion-neutrino chiral symmetric models, in which the primordial
chiral symmetry is broken either explicitly or spontaneously. We
explain how the axion-neutrino interactions may thermalize these
particles in the temperature range $T\sim 1\,$MeV to $T\sim
0.01\,$eV and we demonstrate that no thermal phase transitions
occur in this system. However, as we show the Higgs portal in the
axion sector in the form of higher order non-renormalizable
operators can induce such a thermal phase transition at so low
temperatures and actually the neutrinos play a fundamental role in
these late-time phase transitions. In section IV we discuss the
qualitative astrophysical and cosmological effects of such
late-time thermal phase transitions, which actually occur in the
redshift range $z\sim 385-37$. Finally, the conclusions follow at
the end of the article.

\section{A Brief Overview of the Misalignment Axion}

To date, no sign of weakly interacting massive particles (WIMPs)
has ever been found. There is however an alternative to WIMPs, the
axion particle which is believed to have a very small mass.
Indeed, the axion is believed to have a mass in the sub-eV region
$10^{-6}-10^{-27}$eV \cite{Marsh:2015xka}. Apart from the QCD
axion, which is rather constrained, there is another appealing
axionic model, the misalignment axion
\cite{Marsh:2015xka,Co:2019jts}, in which the axion emerges as the
angular component of a complex scalar field which has a primordial
broken Peccei-Quinn $U(1)$ symmetry. This symmetry is also
believed to be broken during inflation, and the axion begins
misaligned its evolution to the potential minimum, with a large
initial vacuum expectation value $\phi_i\sim f_a$, with $f_a$
being the axion decay constant, which is of the order
$f_a>10^{9}\,$GeV. The axion potential is,
\begin{equation}\label{axionpotentnialfull}
V_a(\phi )=m_a^2f_a^2\left(1-\cos \left(\frac{\phi}{f_a}\right)
\right)\, ,
\end{equation}
and as the axion rolls to its minimum, it holds true that
$\phi/f_a<1$, thus the axion potential behaves as,
\begin{equation}\label{axionpotential}
V_a(\phi )\simeq \frac{1}{2}m_a^2\phi^2\, .
\end{equation}
The misalignment axion has two distinct versions, the canonical
misalignment \cite{Marsh:2015xka}, in which the axion rolls with
zero initial kinetic energy and the kinetic misalignment axion
\cite{Co:2019jts} in which the axion rolls with non-zero kinetic
energy. Regardless of the model, when the Hubble rate becomes of
the order of the axion mass, the axion starts to oscillate around
its potential minimum and redshifts as cold dark matter. The
difference between the two models is that in the case of the
kinetic axion, the axion oscillations commence at a later time,
deeply in the reheating era. In this work we shall provide a
mechanism of generation for the cosine axion potential and we
shall explore how the axion oscillations may be disrupted and this
may lead to a late-time phase transition.

\section{Late-time Phase Transitions with Axion-neutrino Couplings and Higgs-axion Higher Order Operators}

In this paper we shall consider the possibility that late-time
phase transitions occur only due to the existence of
axion-neutrino couplings. As we will show, this is not possible
when one solely includes the axion scalar contributions to the
neutrino effective potential. In similar works in the literature,
in order to generate late-time phase transitions, only the fermion
contribution to the effective potential was considered, which as
we show, this result is simply wrong. In addition, several
cancellations in the effective potential, using the
$\overline{\mathrm{MS}}$ renormalization scheme, indicate that
$Z_N$-symmetric chiral neutrino-axions Lagrangians do not induce
any late-time phase transitions, as we will show explicitly.
However, if higher-order operators between the axion and Higgs are
combined with neutrino-axion couplings, these may eventually
induce late-time phase transitions, as we will show. In fact, the
late-time phase transitions in the latter case are induced due to
the non-trivial neutrino-axion couplings.

We shall consider two types of phenomenological axion-neutrino
models, each of which has its own advantages. In both the
axion-neutrino models we shall consider, there is a primordial
chiral symmetry in the neutrino sector, which shall be broken
either explicitly or spontaneously. Each of the two models has its
own inherent phenomenological significance as we shall
demonstrate. The phenomenologically more important model, which
contains an explicit chiral symmetry breaking term, containing a
coupling of a single neutrino species $\nu_L,\nu_R$ with the
angular component $\phi$ of a pseudo Nambu-Goldstone boson $\Phi$
originating by the primordial breaking of a $U(1)$ Peccei-Quinn
symmetry, so the Lagrangian of this simple model has the following
form,
\begin{equation}\label{axionneutrinolagraM1}
\mathcal{L}=\frac{1}{2}\partial^{\mu}\phi\partial_{\mu}\phi+i\bar{\nu}_L\gamma^{\mu}\partial_{\mu}\nu_L+i\bar{\nu}_R\gamma^{\mu}\partial_{\mu}\nu_R
+m_0\bar{\nu}_L\nu_Re^{i\frac{\phi}{f}}-\epsilon\bar{\nu}_L\nu_R+\Lambda_a^4+\mathrm{H.C.}\,
,
\end{equation}
where the term which couples the axion with the neutrino may arise
from a direct Yukawa coupling between the neutrino with a complex
scalar field $\Phi$ associated with the spontaneous symmetry
breaking of the primordial $U(1)$ Peccei-Quinn symmetry which gave
rise to the axion. This primordial Yukawa coupling term is of the
form $g\Phi \bar{\nu}_L \nu_R$, and when the complex scalar field
$\Phi$ acquires a non-zero vacuum expectation value of the form
$\langle \Phi \rangle=\frac{1}{\sqrt{2}}fe^{i\frac{\phi}{f}}$, the
term $\sim m_0\bar{\nu}_L\nu_Re^{i\frac{\phi}{f}}$ arises, with
$m_0=\frac{g f}{\sqrt{2}}$. The scale $\Lambda_a$ is
 basically connected with the energy scale at which the spontaneous
breaking of the primordial $U(1)$ Peccei-Quinn symmetry breaking
occurs. We have to note that the above Lagrangian does not include
a coupling of the axion to the photons. This however can be
generated by 1-loop effects from the Weyl anomaly even if it is
absent at tree order, see Ref. \cite{Lambiase:2022ucu} for a
similar procedure. Then in principle such a coupling, in
combination with the axion-neutrino coupling, could potentially
lead to a neutrino magnetic moment at 1-loop level, which is
generally constrained. This effect could indirectly constrain the
axion-neutrino coupling. However this task exceeds the aims and
scopes of the present work, we just discuss it though for
completeness since it is quite intriguing to think this aspect of
axion-neutrino couplings\footnote{see also the text below Eq.
(\ref{fifthforcecouplings})}. Now the important feature of this
model, apart from the interaction between the axion and the
neutrino, is the explicit chiral symmetry breaking term $\sim
\epsilon \nu_L\nu_R$, which breaks the chiral symmetry
primordially. The most important effect of this term and of the
term $\sim m_0\bar{\nu}_L\nu_Re^{i\frac{\phi}{f}}$ is that at
one-loop they induce a quadratic divergence in the Lagrangian of
the form,
\begin{equation}\label{quadratic}
\mathcal{L}^{1-loop}=-\frac{m_0\epsilon\Lambda^2}{8\pi^2}\cos
(\frac{\phi}{f})\, ,
\end{equation}
where $\Lambda$ is a cutoff of the theory which can be of the
order of the axion decay constant $f$ or larger. We can choose
without loss of generality,
\begin{equation}\label{choiceofscale}
\Lambda_a^2=\frac{m_0\epsilon\Lambda^2}{8\pi^2}\, ,
\end{equation}
where $\Lambda_a$ appeared firstly in the Lagrangian
(\ref{axionneutrinolagraM1}), thus the total Lagrangian including
the one-loop correction term reads,
\begin{equation}\label{axionneutrinolagraM1final}
\mathcal{L}^{1-loop}=\frac{1}{2}\partial^{\mu}\phi\partial_{\mu}\phi+i\bar{\nu}_L\gamma^{\mu}\partial_{\mu}\nu_L+i\bar{\nu}_R\gamma^{\mu}\partial_{\mu}\nu_R
+m_0\bar{\nu}_L\nu_Re^{i\frac{\phi}{f}}-\epsilon\bar{\nu}_L\nu_R+\Lambda_a^4\left(
1-\cos(\frac{\phi}{f})\right)+\mathrm{H.C.}\, ,
\end{equation}
therefore, even though the initial Lagrangian did not contain any
mass term for the axion field $\phi$, a mass term is generated at
one-loop due to the presence of the explicit chiral symmetry
breaking term. This is one of the attributes of the present model,
since this explicit chiral symmetry breaking may serve as a
mechanism for generating the axion mass term in the form of a
cosine potential. Such cosine potentials are known to be generated
in a non-perturbative way, see for example \cite{Marsh:2015xka}.
In our case, the explicit breaking of the primordial chiral
symmetry leads to the generation of a cosine mass term for the
axion due to one-loop effects. Now let us proceed to the analysis
of the above model. The induced cosine potential term of the axion
further acts as a spontaneous symmetry breaking term which breaks
the chiral symmetry in the neutrino sector down to a residual
$Z_2$ shift symmetry $\phi\to \phi+n\pi f$, which also further
protects the axion from having extra quadratic corrections in its
mass from 1-loop contributions. Such cosine potentials may arise
in the theory in a non-perturbative way if the associated symmetry
has an anomalous current. We provided a physical way on how this
axion mass term may arise in the theory. Furthermore, the explicit
chiral symmetry breaking term has important implications in the
neutrino sector since it induces a fifth force between neutrinos,
due to induced existence of the derivative couplings of the form
$\sim
\frac{1}{f}\partial^{\mu}\phi\bar{\nu}\gamma_5\gamma_{\mu}\nu$.
Thus the axion $\phi$ will be the mediator of a fifth force in the
neutrino sector. The relative strength of this extra force
coupling constant $G$, compared with the coupling of the Newton
gravity $G_N$, has the following form \cite{Hill:1988bu},
\begin{equation}\label{fifthforcecouplings}
\frac{G}{G_N}\sim \frac{\epsilon^2M_P^2}{m_0^2f^2}\, ,
\end{equation}
where $M_P$ is the reduced Planck mass. At a later point, when we
use phenomenological arguments to determine the values of the free
parameters, we shall also discuss the value of the fifth force
coupling constant. In the absence of the explicit chiral symmetry
breaking term accompanied with CP-violation, the Adler decoupling
is violated, and thus no fifth forces  arise in the neutrino
sector, mediated by the axion. The second model we shall consider
in a later section will exactly deal with this case. Also we shall
not take into account the effects of gauge fields $F$, via the
axial anomaly $\bar{\nu}\gamma_5\gamma_{\mu}\nu=c\,\tilde{F}F$
which if are present in plasma state, can trigger a late-time
phase transition via the axial anomaly term $\sim c\,\phi
\tilde{F}F$ \cite{Frieman:1991tu}. This could explicitly break the
symmetries of the system via loops and can cause late-time phase
transitions \cite{Frieman:1991tu}, but we shall not take into
account gauge fields at all in this letter. Diagonalizing the
fermion sector, the effective Lagrangian of the axion-neutrino
sector including the one-loop corrections reads,
\begin{equation}\label{axionneutrinolagraM1finaldiag}
\mathcal{L}^{1-loop}=\frac{1}{2}\partial^{\mu}\phi\partial_{\mu}\phi+i\bar{\nu}\slashed{\partial}\nu-\frac{1}{2}\partial^{\mu}\phi\bar{\nu}\gamma_{\mu}\gamma_5\nu
+m_{\nu}\bar{\nu}\nu +\Lambda_a^4\left(
1-\cos(\frac{\phi}{f})\right)\, ,
\end{equation}
where the effective mass for the neutrino reads,
\begin{equation}\label{effectiveneutrinomass}
m^2_{\nu}=m_0^2+\epsilon^2-2m_0\epsilon\cos(\frac{\phi}{f})\, .
\end{equation}
We can easily calculate the finite-temperature corrected effective
potential for the axion-neutrino model, including the zero
temperature one-loop corrections and the tree-order effective
potential of the misalignment axion,
\begin{equation}\label{finitetempeffectivepotM1}
V(\phi)=V_{tree}+V^{1-loop}_{T=0}(\phi)+V^{1-loop}_{T\neq
0}(\phi)\, ,
\end{equation}
where for the tree potential,
\begin{equation}\label{treeM1}
V_{tree}=\frac{m_a^2}{2}\phi^2-\frac{\Lambda_a^4}{24 f^4}\phi^4\,
,
\end{equation}
we considered small displacements of the axion field from the
minimum of the potential ($\frac{\phi}{f}\ll 1$), also
$m_a^2=\Lambda^4/f^2$ and with $V^{1-loop}_{T=0}(\phi)$ being
equal to,
\begin{equation}\label{oneloopzerotM1}
V^{1-loop}_{T=0}(\phi)=\frac{m_{\phi}^4}{64\pi^2}\left(\ln\frac{m_{\phi}^2}{\mu^2}-\frac{3}{2}
\right)-\frac{3}{16}\frac{m_{\nu}^4}{\pi^2}\left(\ln\frac{m_{\nu}^2}{\mu^2}-\frac{3}{2}
\right)\, ,
\end{equation}
with $\mu$ being an arbitrary for the moment renormalization scale
and $m_{\phi}^2=\frac{d^2V_{tree}}{d\phi^2}$. Also the one-loop
finite-temperature correction term $V^{1-loop}_{T\neq 0}(\phi)$
reads,
\begin{equation}\label{oneloopfinitet1M1}
V^{1-loop}_{T\neq
0}(\phi)=\frac{m_{\nu}T^2}{4}+\frac{3}{16\pi^2}m_{\nu}^4\ln\left(\frac{m_{\nu}^2}{a_fT^2}
\right)+\frac{m_{\phi}^2T^2}{24}-\frac{1}{12\pi}m_{\phi}^3T-\frac{m_{\phi}^4}{64\pi^2}\ln\left(\frac{m_{\phi}^2}{a_b
T^2} \right)\, ,
\end{equation}
with $\ln(a_b)=5.4076$ and $\ln(a_f)=2.6351$. Notice the
cancellation between the mass-dependent one-loop terms at finite
temperature and zero temperature for both the neutrino and the
axion, therefore, the resulting effective potential reads,
\begin{equation}\label{finitetempeffectivepotfinalM1}
V(\phi)=\frac{m_a^2}{2}\phi^2-\frac{\Lambda_a^4}{24
f^4}\phi^4+\frac{m_{\phi}^2T^2}{24}-\frac{1}{12\pi}m_{\phi}^3T+\frac{m_{\nu}T^2}{4}+\frac{m_{\phi}^4}{64\pi^2}\left(\ln\left(\frac{a_b
T^2}{\mu^2} \right)
-\frac{3}{2}\right)-\frac{3m_{\nu}^4}{16\pi^2}\left(\ln\left(\frac{a_f
T^2}{\mu^2} \right) -\frac{3}{2}\right) \, .
\end{equation}
Now let us discuss an important issue having to do with the
thermalization of the axion, before proceeding to the study of the
finite temperature effective potential. The neutrino decouples
from the electroweak thermal bath at a temperature $T_D\sim
1\,$MeV. On the other hand, the only way that the axion is
thermalized is via its couplings with the neutrinos and recall
that the Yukawa coupling which controls this interaction has the
form $g\Phi \bar{\nu}_L \nu_R$, with $\langle \Phi
\rangle=\frac{1}{\sqrt{2}}fe^{i\frac{\phi}{f}}$. The condition
that the axion field $\phi$ is at thermal equilibrium with the
neutrino bath is $T\geq \frac{f^2}{M_P}$ so for an axion decay
constant of the order $f\sim 10^9\,$GeV, the temperature above
which the axion can be considered at thermal equilibrium with the
axion background is approximately beyond $T_c\sim 10^9\,$eV. Thus
the axion decouples thermally from its background via decay rates
quite early in the Universe's evolution, if such high reheating
temperature was achieved. Even in the case that the reheating
temperature was not so high, the axion would never thermalize via
its interactions with the neutrino. However,  we can consider that
the thermal equilibrium between the axion and the neutrino occurs
in a classical way between ensembles, identical to the thermal
equilibrium considered in Ref. \cite{Frieman:1991tu}. So in a way
the axion-neutrino thermal equilibrium is not due small
interaction rates between axion and neutrinos with given
incoherent scattering between neutrinos and axions, but it is by
considering the neutrinos and axions as classical coherent fields
and classical ensembles. One can consider such interaction as a
collective classical interaction of the coherent axion
oscillations with the neutrino thermal bath. Effectively, the
temperature corrections are justified by calculating values of
operators to which the axion couples, such as $\bar{\nu}\nu$, in
an appropriate density matrix for neutrinos\footnote{See Ref.
\cite{Frieman:1991tu} page 1230 above Eq. (2.20).}. Using the
argument of Ref. \cite{Frieman:1991tu}, this thermal interaction,
without taking into account the microphysical decay rates between
axions and neutrinos, is similar in the way that a classical
massive body can feel the gravity of another massive object, that
is, it is not of importance to consider the reaction rate of
gravitons on baryons. Thus the thermal equilibrium of the axion
with the neutrino background is justified above $T_c\sim
0.01\,$eV, at which temperature the neutrino decouples. Even if
someone considers that the axion would be thermalized with the
Standard Model particles, by particle interactions with the
neutrino, for a reheating temperature as high as 1000GeV this
would not be true. To have an idea on this, the axion-neutrino
interactions that keep the axion in thermal equilibrium have rates
of the order $\Gamma \sim \frac{T^3}{f^2}$ hence for $T\sim
1000\,$GeV, the rate of the reaction is $\Gamma\sim
\mathcal{O}(1)\,$eV. On the other hand, the cross sections of the
weak interactions beyond the neutrino decoupling temperature are
of the order $\sigma\sim \frac{G_F^2\,s}{\pi}$ with
$G_F=\frac{g_w^2}{8m_w^2}$ and $s=(2E)^2$ where $E$ is the center
of mass energy of the particles participating in the interaction.
Typical electroweak interactions in which the neutrino
participates are $W^-\to e^-\nu_e$ and $Z\to \nu_e\bar{\nu}_e$,
with the first having a rate of the order $\Gamma(W^-\to
e^-\nu_e)=2.1\,$GeV and a branching ratio of the order $BR(W^-\to
e^-\nu_e)=67\%$ while the second has a rate $\Gamma(Z\to
\nu_e\bar{\nu}_e)=167\,$MeV and a branching ratio $BR(Z\to
\nu_e\bar{\nu}_e)=21\%$. Apparently, it is by far more likely for
a neutrino to participate to an electroweak interaction beyond the
neutrino decoupling temperature, compared to the neutrino
participation in an axion-neutrino interaction. Finally, it would
be important to justify the thermal bath constituted by neutrinos
in the post neutrino decoupling epoch. In the post neutrino
decoupling epoch, the neutrino distribution function is described
by that of a particle at thermal equilibrium, at least when the
temperature is still larger that the neutrino mass
\cite{Frieman:1991tu}, with the effective neutrino thermal bath
temperature $T$ at the cosmic time instance $t$ being $T\sim
\frac{a(t_D)T_D}{a(t)}$, where $a(t)$ is the scale factor, $T_D$
the decoupling temperature, and $t_D$ is the cosmic time instance
at the neutrino decoupling. Of course it is conceivable that the
axion decouples from the thermal equilibrium when the temperature
becomes of the order of the neutrino mass $m_{0}$, since the
thermal equilibrium at $T\sim m_{0}$ is abruptly disrupted.

Now let us proceed to the analysis of the effective potential at
finite temperature given in Eq.
(\ref{finitetempeffectivepotfinalM1}). We shall consider an
neutrino absolute rest mass of the order $m_0\sim 0.01\,$eV. This
choice is compatible with current constraints on the rest mass of
neutrinos, which come from both experimental evidence and
theoretical constraints. The observational constraints on the
absolute rest mass of neutrinos come from cosmological
observations based on the CMB, and the Lyman-$\alpha$ forest.
According to the latest Planck data the sum of the masses of the
three neutrinos is constrained to be $\sum_{i}^3
m_{m_{\nu_i}}<0.12$ at 95$\,\%\,$CL \cite{Planck:2018vyg} in the
absence of sterile neutrinos.

\begin{figure}[h!]
\centering
\includegraphics[width=20pc]{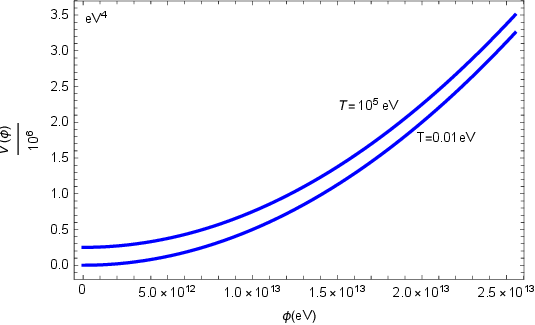}
\caption{The effective potential (units in eV$^4$)  of the
axion-neutrino thermal bath, for the model of Eq.
(\ref{axionneutrinolagraM1finaldiag}), below $T_D\sim 1\,$MeV and
down to $T\sim 0.01\,$eV, for two characteristic values of the
temperature. No sign of a thermal phase transition exists for this
simple interacting system.} \label{plot1M1}
\end{figure}

Now let us use some phenomenological arguments in order to
determine the value of $\epsilon$. So let us assume that the axion
has a mass of the order $m_a\sim 10^{-10}\,$eV which is highly
motivated by recent studies which predict such a mass for the
axion, and are based on relative Gamma Ray Burst observations
\cite{Hoof:2022xbe,Li:2022pqa}. Also in the context of the present
model $\Lambda_a^4=m_a^2 f^2$, thus we choose the renormalization
scale $\Lambda$ to be $\Lambda=10^9\,$GeV which is of the same
order as the axion decay constant. By using these assumptions, it
follows that $\epsilon\sim \mathcal{O}(10^{-7})\,$GeV. For these
values for the free parameters, the relative strength of the fifth
force, compared to Newton's gravity constant, given in Eq.
(\ref{fifthforcecouplings}), is of the order $\frac{G}{G_N}\sim
10^{-10}$, which is well compatible with the current constraints
on the fifth force. Let us study now numerically the behavior of
the effective potential at finite temperature. For the plots, we
shall also assume that $\phi\ll f$, hence the effective mass of
the axion at tree order is $m_{eff}^2=m_a^2-\frac{\phi^2}{24
f^4}$. The effective potential at $T\sim 0.01\,$eV and $T\sim
10^5\,$eV is plotted in Fig. \ref{plot1M1}.
As it can be seen in Fig. \ref{plot1M1}, no phase transition
occurs. Indeed, the axion-neutrino effective potential has only
one minimum at the origin, both at zero temperature and at high
temperature. Thus in effect, the axion oscillations are not
disturbed, even though the axion gets thermalized with the
neutrino bath. This result is in contrast to Ref.
\cite{Frieman:1991tu}, and now let us demonstrate the difference
between our result and that of Ref. \cite{Frieman:1991tu}. The
problem between our approach and that of Ref.
\cite{Frieman:1991tu} is that in practise, the mass dependent
logarithmic contributions between the zero temperature one-loop
fermion and boson contributions are cancelled with the
finite-temperature contributions, and specifically the
cancellation occurs between the terms containing
$\frac{m_{\phi}^4}{64\pi^2}\ln \left(m_{\phi}^2\right)$ for the
boson part and $\frac{12m_{\nu}^4}{64\pi^2}\ln
\left(m_{\nu}^2\right)$ for the neutrino contribution. In Ref.
\cite{Frieman:1991tu}, this cancellation is not taken into account
and thus it seems that the result is not valid, plus no other
temperature dependent-mass containing terms are also not taken
into account. We believe that our result is transparent, and no
phase transition occurs at late-times for the thermalized
axion-neutrino system. Let us note here that the era in which a
possible phase transition is checked in the case at hand is in the
temperature range $T\sim(0.1,0.01)\,$eV so basically in the
redshift range $z\sim (385,37)$. Hence, if a phase transition
occurs at late times, it has to occur when the temperature drops
well below the neutrino decoupling temperature. The critical
temperature should in principle be of the order of the neutrino
mass, $m_{0}\sim 0.01$ so $T\sim 0.01$eV so it is expected that
any possible phase transition will take place somewhere at the
redshift interval $z\sim (385,37)$.

The absence of finite temperature phase transitions persists even
if we consider a generalization of the above axion-neutrino
system, considering $N$ different fermion flavors, with a $Z_N$
symmetric Lagrangian \cite{Frieman:1991tu},
\begin{align}\label{z3symmetriclagrangian}
&\mathcal{L}=\frac{1}{2}\partial_{\mu}\phi\partial^{\mu}\phi+\sum_{j=0}^{N-1}\left(i\bar{\nu}_j\gamma^{\mu}\partial_{\mu}\nu_j+\left(m_0+\epsilon^{i(\frac{\phi}{f}+\frac{2\pi
j}{N})} \right)\bar{\nu}_{jL}\nu_{jR}\right)+
\Lambda_a^4\left(1-\cos(\frac{\phi}{f}) \right)\, .
\end{align}
The explicit chiral symmetry breaking terms generate quadratic
divergences at one-loop level, so the effective Lagrangian further
contains the following terms,
\begin{equation}\label{effectiveznterms}
\sum_{j=0}^{N-1}\frac{m_0\epsilon\Lambda^2}{8\pi^2}\cos\left(
\frac{\phi}{f}+\frac{2p j}{N}\right)\, .
\end{equation}
the fermion masses in this case are,
\begin{equation}\label{fermionmasszn}
M_j^2=m_0^2+\epsilon^2+2m_0\epsilon \cos\left(
\frac{\phi}{f}+\frac{2p j}{N}\right)\, ,
\end{equation}
and due to the $Z_N$ symmetry we have the following relations that
hold true,
\begin{equation}\label{relation1}
\sum_{j=0}^{N-1}M_j^2=N(m_0^2+\epsilon^2)\, ,
\end{equation}
\begin{equation}\label{relation2}
\sum_{j=0}^{N-1}M_j^4=N\left(2m_0^2\epsilon^2+(m_0^2+\epsilon^2)^2\right)\,
.
\end{equation}
In the $\bar{\mathrm{MS}}$ renormalization scheme, the one-loop
finite temperature corrected axion-neutrino effective potential
reads,
\begin{align}\label{finitetemperatureaxionneutrinoexplicit1}
V(\phi)&=\Lambda_a^4\left(1-\cos(\frac{\phi}{f})
\right)+\frac{m_{\phi}^4}{64\pi^2}\left(
\frac{a_b\,T^2}{\mu^2}-\frac{3}{2}\right)+\frac{m_{\phi}^2T^2}{24}-\frac{1}{12\pi}\left(
m_{\phi}^{3/2}\right)T\\
\notag &
-\sum_{j=0}^{N-1}\frac{3}{16}\frac{M_j^4}{\pi^4}\left(\log
\left(\frac{M_j^2}{\mu^2} \right)-\frac{3}{2}
\right)+\frac{3}{16\pi^2}\sum_{j=0}^{N-1}M_j^4\log\left(\frac{M_j^2}{a_f\,T^2}\right)+\sum_{j=0}^{N-1}M_j^2\frac{T^2}{4}\,
,
\end{align}
where $\mu$ is the renormalization scale, and due to the fact
that,
\begin{align}\label{neweefinter}
&-\sum_{j=0}^{N-1}\frac{3}{16}\frac{M_j^4}{\pi^4}\left(\log
\left(\frac{M_j^2}{\mu^2} \right)-\frac{3}{2}
\right)+\frac{3}{16\pi^2}\sum_{j=0}^{N-1}M_j^4\log\left(\frac{M_j^2}{a_f\,T^2}\right)=\\
\notag &
\sum_{j=0}^{N-1}\frac{3}{16}M_j^4\frac{3}{2}-\sum_{j=0}^{N-1}\frac{M_j^4}{\pi^2}\left(\log
M_j^2 -\log
\mu^2\right)+\frac{3}{16\pi^2}\sum_{j=0}^{N-1}M_j^4\left(\log
M_j^4 -\log\left(a_f T^2 \right)\right)\, ,
\end{align}
the last two terms in the effective potential
(\ref{finitetemperatureaxionneutrinoexplicit1}) can be written as
follows,
\begin{equation}\label{extraenteka}
-\sum_{j=0}^{N-1}\frac{3}{16}\frac{M_j^4}{\pi^4}\left(\log
\left(\frac{a_f\,T^2}{\mu^2} \right)-\frac{3}{2}
\right)+\sum_{j=0}^{N-1}M_j^2\frac{T^2}{4}\, .
\end{equation}
Due to relations (\ref{relation1}) and (\ref{relation2}), the two
terms above can be written as,
(\ref{finitetemperatureaxionneutrinoexplicit1}),
\begin{equation}\label{extrearelation1}
-\sum_{j=0}^{N-1}\frac{3}{16}\frac{M_j^4}{\pi^4}\left(\log
\left(\frac{a_f\,T^2}{\mu^2} \right)-\frac{3}{2}
\right)=N\left(2m_0^2\epsilon^2+(m_0^2+\epsilon^2)^2\right)\left(\log
\left(\frac{a_f\,T^2}{\mu^2} \right)-\frac{3}{2} \right)\, ,
\end{equation}
and
\begin{equation}\label{extrearelation2}
\sum_{j=0}^{N-1}M_j^2\frac{T^2}{4}=\frac{N}{4}\left(m_0^2+\epsilon^2
\right)T^2\, ,
\end{equation}
which are both field independent. This feature can also be seen by
adding the relevant terms in Ref. \cite{Frieman:1991tu}, so
basically the result of Ref. \cite{Frieman:1991tu} cannot be
correct, because in the fermion sector there is no field dependent
term. Thus only the axion contributes to the $Z_N$ symmetric
axion-neutrino system, and it can easily be shown that there is no
phase transition regardless the value of the temperature, which is
in contrast with the result of Ref. \cite{Frieman:1991tu}. The
discrepancy occurs due to the cancellation of the field dependent
contributions in the effective potential, which we demonstrated
above, which can also be observed in Ref.  \cite{Frieman:1991tu}
but is not taken into account.

Now let us consider the axion and a single neutrino sector in the
presence of a spontaneous breaking term of the chiral symmetry in
the form $\sim \cos\left( \frac{\phi}{f}\right)$, in the absence
of an explicit breaking term. The simple chiral symmetric
axion-neutrino Lagrangian with a single neutrino species
containing the cosine spontaneous chiral symmetry breaking
potential for the axion scalar is,
\begin{equation}\label{axionneutrinolagra}
\mathcal{L}=\frac{1}{2}\partial^{\mu}\phi\partial_{\mu}\phi+i\bar{\nu}_L\gamma^{\mu}\partial_{\mu}\nu_L+i\bar{\nu}_R\gamma^{\mu}\partial_{\mu}\nu_R
+m_0\bar{\nu}_L\nu_Re^{i\frac{\phi}{f}}+\Lambda_a\left(1-\cos\left(
\frac{\phi}{f}\right) \right)+\mathrm{H.C.}\, ,
\end{equation}
where in the same way as in the previous models, the term which
couples the axion field with the neutrino may arise from a direct
Yukawa coupling between the complex scalar field $\Phi$ associated
with the spontaneous symmetry breaking of the primordial $U(1)$
Peccei-Quinn symmetry which gave rise to the axion. In the present
case, the scale $\Lambda_a$ is basically connected with the energy
scale at which the spontaneous breaking of the primordial $U(1)$
Peccei-Quinn symmetry breaking occurs. The cosine potential term
of the axion breaks the chiral symmetry in the neutrino sector
down to a residual $Z_2$ shift symmetry $\phi\to \phi+n\pi f$,
which also protects the axion from having quadratic corrections in
its mass from 1-loop contributions. Thus in the present case, the
axion receives its mass by an undetermined non-perturbative
mechanism and the residual $Z_2$ shift symmetry of the cosine
potential protects the axion from receiving corrections to its
mass due to one loop quadratic divergences. Such cosine potentials
may arise in the theory in a non-perturbative way if the
associated symmetry has an anomalous current. For the purposes of
this letter, we shall assume that this cosine potential term does
not arise from a fundamental theoretical process, but it is of
unknown origin, or some non-perturbative arguments of the
underlying theory give rise to this term. Also in the present
context, one has not to take into account fifth forces in the
neutrino sector, due to induced existence of the derivative
couplings of the form $\sim
\frac{1}{f}\partial^{\mu}\phi\bar{\nu}\gamma_5\gamma_{\mu}\nu$,
because for small $\phi$ momentum, the $\phi$ emission and
absorption amplitudes will tend to zero \cite{Frieman:1991tu}. Due
to this decoupling procedure, the axion $\phi$ will not be the
mediator of a fifth force in the neutrino sector. However, as we
showed in the previous models, in the presence of an explicit
chiral symmetry breaking term accompanied with CP-violation, the
Adler decoupling we discussed is violated, and thus fifth forces
may arise in the neutrino sector, mediated by the axion. We can
easily calculate the finite-temperature corrected effective
potential for the axion-neutrino model, including the zero
temperature one-loop corrections and the tree-order effective
potential of the misalignment axion,
\begin{equation}\label{finitetempeffectivepot}
V(\phi)=V_{tree}+V^{1-loop}_{T=0}(\phi)+V^{1-loop}_{T\neq
0}(\phi)\, ,
\end{equation}
where for the tree potential,
\begin{equation}\label{tree}
V_{tree}=\frac{m_a^2}{2}\phi^2-\frac{\Lambda_a^4}{24 f^4}\phi^4\,
,
\end{equation}
we again considered small displacements of the axion field from
the minimum of the potential ($\frac{\phi}{f}\ll 1$), which are
valid for the misalignment axion models. Also
$m_a^2=\Lambda^4/f^2$ and with $V^{1-loop}_{T=0}(\phi)$ being
equal to,
\begin{equation}\label{oneloopzerot}
V^{1-loop}_{T=0}(\phi)=\frac{m_{\phi}^4}{64\pi^2}\left(\ln\frac{m_{\phi}^2}{\mu^2}-\frac{3}{2}
\right)-\frac{3}{16}\frac{m_0^4}{\pi^2}\left(\ln\frac{m_{0}^2}{\mu^2}-\frac{3}{2}
\right)\, ,
\end{equation}
with $\mu$ being again an arbitrary for the moment renormalization
scale and $m_{\phi}^2=\frac{d^2V_{tree}}{d\phi^2}$. Also the
one-loop finite-temperature correction term $V^{1-loop}_{T\neq
0}(\phi)$ reads,
\begin{equation}\label{oneloopfinitet1}
V^{1-loop}_{T\neq
0}(\phi)=\frac{m_0T^2}{4}+\frac{3}{16\pi^2}m_0^4\ln\left(\frac{m_0^2}{a_fT^2}
\right)+\frac{m_{\phi}^2T^2}{24}-\frac{1}{12\pi}m_{\phi}^3T-\frac{m_{\phi}^4}{64\pi^2}\ln\left(\frac{m_{\phi}^2}{a_b
T^2} \right)\, ,
\end{equation}
and recall that $\ln(a_b)=5.4076$ and $\ln(a_f)=2.6351$. Notice
again the cancellation between the mass-dependent one-loop terms
at finite temperature and zero temperature for both the neutrino
and the axion fields, therefore, the resulting effective potential
reads,
\begin{equation}\label{finitetempeffectivepotfinal}
V(\phi)=\frac{m_a^2}{2}\phi^2-\frac{\Lambda_a^4}{24
f^4}\phi^4+\frac{m_{\phi}^2T^2}{24}-\frac{1}{12\pi}m_{\phi}^3T+\frac{m_0T^2}{4}+\frac{m_{\phi}^4}{64\pi^2}\left(\ln\left(\frac{a_b
T^2}{\mu^2} \right)
-\frac{3}{2}\right)-\frac{3m_0^4}{16\pi^2}\left(\ln\left(\frac{a_f
T^2}{\mu^2} \right) -\frac{3}{2}\right) \, .
\end{equation}
The same thermalization arguments for the neutrino-axion system
hold true, as in the previous models thus the temperature above
which the axion can be at thermal equilibrium with the axion
background is beyond $T_c\sim 0.01\,$eV and below the neutrino
electroweak decoupling temperature $T_D\sim 1\,$MeV, above which
the neutrino is at thermal equilibrium with the particles which
interact with it via the weak interactions.

\begin{figure}[h!]
\centering
\includegraphics[width=19pc]{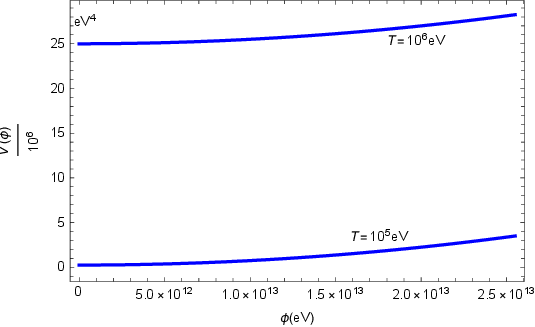}
\includegraphics[width=20pc]{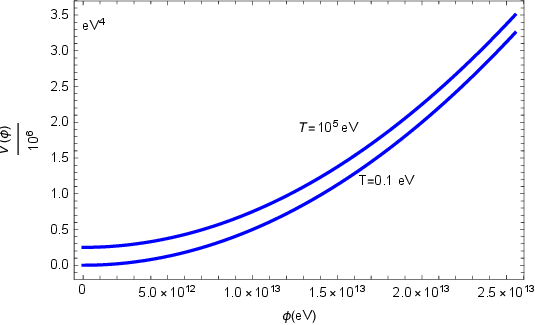}
\caption{The effective potential (units in eV$^4$) of the
axion-neutrino thermal bath below $T_D\sim 1\,$MeV and up to
$T\sim 0.1\,$eV, for three characteristic values of the
temperature. No sign of a thermal phase transition exists for this
simple interacting system.} \label{plot1}
\end{figure}

Also in the context of the present model $\Lambda_a=\sqrt{m_a f}$.
For the plots, we shall also assume that $\phi\ll f$, hence the
effective mass of the axion at tree order is
$m_{eff}^2=m_a^2-\frac{\phi^2}{24 f^4}$. The effective potential
at $T\sim 0.1\,$eV, $T\sim 10^3\,$eV and $T\sim 10^5\,$eV is
plotted in Fig. \ref{plot1}.
As in the previous cases, it can be seen in Fig. \ref{plot1}, no
phase transition occurs as the axion-neutrino effective potential
has only one minimum at the origin, a situation which occurs both
at zero temperature and at high temperature. Thus in effect, the
axion oscillations are not disturbed in this case too, regardless
the axion thermalization with the neutrino bath. The attribute of
the first model we presented in this section, compared to the
second model is that the first model respects the naturalness
argument. Recall that by naturalness it is meant that, the mass
scales of all particles participating in a Lagrangian must not be
fined tuned, but instead must appear as a inherent to the theory
consequence of some mechanism. As t'Hooft also claims, a parameter
is considered naturally small if in its limit to zero, the
symmetry structure of the Lagrangian is enhanced. In such a case
the parameter will in general be multiplicatively renormalizable
and can remain small at all orders in perturbation theory. Also,
in a more restrictive way, in the strong naturalness case, the
mass of ultralight particles must emerge on some symmetry breaking
ground and not being fixed by hand. In the case at hand, the first
model provides a mechanism for the generation of the axion mass,
via an explicit chiral symmetry breaking term, so it is more
complete from a phenomenological point of view, compared to the
second model.

\section{Late-time Phase Transitions Caused by the Higgs Portal Interactions of the Axion}

\begin{figure}
\centering
\includegraphics[width=19pc]{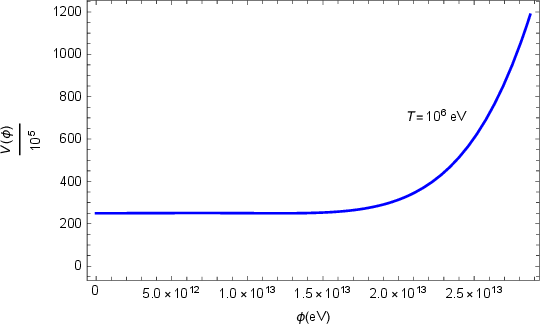}
\includegraphics[width=20pc]{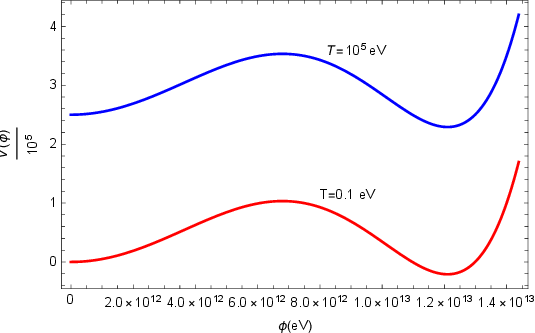}
\caption{The effective potential (units in eV$^4$)  of the
axion-neutrino thermal bath below $T_D\sim 1\,$MeV and down to
$T\sim 0.01\,$eV, for three characteristic values of the
temperature including the effects of a Higgs-axion coupling via
higher order non-renormalizable operators. There is a clear
indication of a first order phase transition in the system, which
is controlled by the fermionic contribution.}\label{plot2}
\end{figure}

Now let us consider an alternative scenario in which the Higgs
portal affects the axion via higher order non-renormalizable
operators. Such a scenario was recently considered in Ref.
\cite{Oikonomou:2023bah}. Specifically we shall assume that the
axion is interacting with the Higgs particle via dimension six and
dimension eight non-renormalizable operators. We assume that the
Universe experiences the standard reheating era epoch and that the
electroweak breaking occurs during the radiation domination era.
Also we assume that the Universe reached a large reheating
temperature, which is larger than $T\sim 100\,$GeV. When the
temperature drops below $T\sim 100\,$GeV, the electroweak breaking
occurs and thus the Higgs particle acquires a non-zero vacuum
expectation value and via the Yukawa couplings and other
interactions, gives mass to the Standard Model particles. With
regard to the axion sector, without for the moment taking into
account the neutrino sector, the axion potential has the form,
\begin{equation}\label{axionpotentnialfull}
V_a(\phi )=m_a^2f_a^2\left(1-\cos \left(\frac{\phi}{f_a}\right)
\right)\, ,
\end{equation}
which when $\phi/f_a<1$, can be approximated as,
\begin{equation}\label{axionpotential}
V_a(\phi )\simeq \frac{1}{2}m_a^2\phi^2\, .
\end{equation}
When considering the axion-neutrino sector, such a potential may
arise in the way we described in the previous section, thus by a
spontaneous symmetry breaking term of the chiral symmetry in the
neutrino sector, or via an explicit breaking term of the chiral
symmetry in the neutrino sector, which induces an axion mass via a
one-loop quadratic divergence. Returning back to the Higgs-axion
interaction, the dimension six and dimension eight
non-renormalizable operators of the Higgs to the axion scalar are
of the form \cite{Oikonomou:2023bah} $\sim
\lambda\frac{|H|^2\phi^4}{M^2}$ and $\sim
g\frac{|H|^2\phi^6}{M^4}$, and thus the axion-Higgs effective
potential at tree order is,
\begin{equation}\label{axioneightsixpotential}
V(\phi,h)=V_a(\phi)-m_H^2|H|^2+\lambda_H|H|^4-\lambda\frac{|H|^2\phi^4}{M^2}+g\frac{|H|^2\phi^6}{M^4}\,
,
\end{equation}
with $V_a(\phi)$ appearing in Eq. (\ref{axionpotentnialfull}). The
Higgs scalar before the electroweak breaking has the form
$H=\frac{h+i h_1}{\sqrt{2}}$ and the Higgs particle mass is
$m_H=125$ GeV, while the Higgs self-coupling is defined through
the relation
$\frac{v}{\sqrt{2}}=\left(\frac{-m_H^2}{\lambda_H}\right)^{\frac{1}{2}}$,
with $v$ being the electroweak symmetry breaking scale which is
$v\simeq246\,$GeV. The scale $M$ denotes the scale at which the
effective field theory scale of the non-renormalizable dimension
six and dimensions eight operators at which they originate from
and are active. We shall assume that this effective scale $M$ is
way higher than the electroweak breaking scale, and specifically
we assume that $M$ is of the order $M=20-100\,$TeV. This choice is
not accidental and is highly motivated by the fact that no
particle has ever been observed at the LHC beyond the electroweak
scale and up to energies of the order $\sim 15\,$TeV
center-of-mass. Regarding the parameters $\lambda$ and $g$ these
are the Wilson coefficients of the higher order effective field
theory of the Higgs-axion sector, which will be assumed to be of
the order $\lambda\sim \mathcal{O}(10^{-20})$ and $g\sim
\mathcal{O}(10^{-5})$ for phenomenological reasons, see Ref.
\cite{Oikonomou:2023bah}. After the electroweak breaking, the
standard thermal history in the Standard Model sector occurs, with
a first order phase transition taking place. We shall mainly be
interested in the axion sector and for temperatures below the
neutrino decoupling sector. As we shall see, the presence of
neutrinos accompanied with the higher order Higgs-axion
interactions can cause a physically interesting situation, with a
late-time phase transition occurring. The couplings of the Higgs
to the axion do not affect the electroweak sector, the operators
are non-renormalizable and thus the effects are significantly
suppressed, but these operators can affect the late-time
axion-neutrino system, causing a late-time phase transition during
the post neutrino decoupling era. After the electroweak breaking,
the Higgs particle obtains a vacuum expectation value, thus
$H=v+\frac{h+i h_1}{\sqrt{2}}$, and therefore the higher
dimensional operators are affected. Therefore, in the post
electroweak breaking era, the axion effective potential
$\mathcal{V}_a(\phi)$ becomes,
\begin{equation}\label{axioneffective68}
\mathcal{V}_a(\phi)=V_a(\phi)-\lambda\frac{v^2\phi^4}{M^2}+g\frac{v^2\phi^6}{M^6}\,
,
\end{equation}
where we defined $V_a(\phi)$ in Eq. (\ref{axionpotentnialfull}).
The behavior of the axion during the radiation domination era in
this framework was studied in Ref. \cite{Oikonomou:2023bah}, but
in this work we shall focus on the post neutrino decoupling era.
Thus, let us write in a compact way the current theory, in which
case the Lagrangian before the electroweak breaking is,
\begin{align}\label{axionneutrinolagrawithHiggs}
&\mathcal{L}=\frac{1}{2}\partial^{\mu}H^+\partial_{\mu}H-m_H^2|H|^2+\lambda_H|H|^4-\lambda\frac{|H|^2\phi^4}{M^2}+g\frac{|H|^2\phi^6}{M^4}&
\\ \notag &+\frac{1}{2}\partial^{\mu}\phi\partial_{\mu}\phi+i\bar{\nu}_L\gamma^{\mu}\partial_{\mu}\nu_L+i\bar{\nu}_R\gamma^{\mu}\partial_{\mu}\nu_R
+m_0\bar{\nu}_L\nu_Re^{i\frac{\phi}{f}}+\Lambda_a\left(1-\cos\left(
\frac{\phi}{f}\right) \right)+\mathrm{H.C.}\, ,
\end{align}
thus the tree order effective potential of the axion and neutrino
system, during the post electroweak epoch is,
\begin{equation}\label{treehiggs}
\mathcal{V}_a(\phi)=\frac{m_a^2}{2}\phi^2-\frac{\Lambda_a^4}{24
f^4}\phi^4-\lambda\frac{v^2\phi^4}{M^2}+g\frac{v^2\phi^6}{M^6}\, ,
\end{equation}
and the tree order effective mass for the axion is
$m_{eff}^2(\phi)$,
\begin{equation}\label{axioneffectivemass}
m_{eff}^2(\phi)=\frac{\partial^2 V(\phi,h)}{\partial
\phi^2}=m_a^2-\frac{6 \lambda v^2 \phi^2}{M^2}+\frac{15 g v^2
\phi^4}{M^4}\, ,
\end{equation}
which is derived from the second derivative of the tree order
axion effective potential with respect to the axion field, that is
$m_{eff}^2(\phi)=\frac{\partial^2 \mathcal{V}_a(\phi)}{\partial
\phi^2}$. Including the zero and finite temperature contributions
to the effective potential for the axion-neutrino system, the
final form of the effective potential is in the high temperature
limit,
\begin{align}\label{newaxionpotentialhiggs}
&V_a^{1-loop T\neq 0}=\frac{m_a^2}{2}\phi^2-\frac{\Lambda_a^4}{24
f^4}\phi^4-\lambda\frac{v^2\phi^4}{M^2}+g\frac{v^2\phi^6}{M^6}+\frac{m_{eff}^2(\phi)}{24}T^2
\\ \notag &
-\frac{T}{12\pi}\left(m_{eff}^2(\phi)\right)^{3/2}+\frac{m_{eff}^4(\phi)}{64\pi^2}
\left(\ln a_b-\frac{3}{2} \right)+
+\frac{m_{eff}^4(\phi)}{64\pi^2} \ln\left(
\frac{T^2}{\mu^2}\right)+\frac{m_0\,T^2}{4}-\frac{3\,m_0^4}{16\pi^2}\left(\ln\left(\frac{a_f
T^2}{\mu^2} \right) -\frac{3}{2}\right)\, .
\end{align}
It can also be observed in this case that the logarithmic mass
terms proportional to $m_{eff}^4(\phi)$ and $m_0^4$, both cancel
in the $\bar{\mathrm{MS}}$ renormalization scheme, for both the
fermion and boson sectors. The form of the potential in Eq.
(\ref{newaxionpotentialhiggs}) is valid only in the high
temperature limit, when $\frac{m_{eff}^2(\phi)}{T^2}\ll 1$ and
$\frac{m_0}{T} \ll 1$. As we discussed thoroughly in the previous
section, the axion and the neutrino can be considered at a
coherent classical thermal equilibrium between ensembles of
particles, beyond $T\sim m_{\nu}$. The thermal equilibrium between
the axion and the neutrino continues until the temperature becomes
of the order of the neutrino mass, at which point the neutrino
decouples from the thermal equilibrium. Therefore, for the axion
and the neutrino are considered to be at a coherent thermal
equilibrium for $T\sim >0.01\,$eV, and the phase transition occurs
approximately at $T\sim m_0\sim 0.01\,$eV, which we cannot
approach in a perturbative way. In Fig. \ref{plot2} we plot the
behavior of the effective potential at high and low temperatures,
in the temperature range $T\sim 10^6-0.01\,$eV, using three
characteristic temperatures $T=10^6\,$eV, $T=10^5\,$eV and
$T=0.1\,$eV. As it can be seen in Fig. \ref{plot2}, the potential
at high temperature, which is very close to the era when the
neutrino decouples from the electroweak thermal bath, has one
minimum at the origin, and this picture continues to hold true
until approximately $T\sim 10^5\,$eV where a second minimum is
developed, with the two minima being separated by a potential
barrier. The same physical picture continues to hold true until
the temperature drops to $T=0.1\,$eV, however at $T\sim 10^5\,$eV
the two minima are equally favorable energetically, but when the
temperature drops to $T=0.1\,$eV, the second minimum away from the
origin becomes more favorable energetically. This physical
behavior points out to one thing only, a first order phase
transition is likely to take place. The exciting thing is that
this first order phase transition will take place at temperatures
of the order $T\sim \mathcal{O}(0.1-0.01)$eV until the temperature
becomes of the order of the neutrino mass, which we assumed to be
of the order $m_{0}\sim 0.01\,$eV. This is a very interesting
physical situation, since this first order phase transition is
actually a late-time phase transition which happens globally to
the Universe at the redshift corresponding to the phase transition
temperature. Now the interesting part is the phase transition
era's redshift range, so if we assume that the temperature range
at which the first order phase transition might take place is
$T\sim 0.1-0.01\,$eV, the redshift range is $z\sim 385-37$, so it
is possible to have a global, literally Universal, first order
phase transition occurring in the era after the recombination,
which occurred at $z\sim 1100$, at which era the CMB originates.
It is vital to note that if the neutrino is absent from the model
we presented in this section, the first order phase transition
does not occur at all.
This is a physically interesting situation from many aspects, and
this scenario tights well with many astrophysical problems and
potential issues, and also with the $H_0$-tension problems as we
will discuss in the next section. In the model at hand, before the
phase transition occurs, the axion is well protected at the origin
of the effective potential, continuing its oscillations, not
destabilized and thus it describes perfectly cold dark matter,
however once the first order phase transition occurs, the axion is
at the second minimum, which compared with the Higgs electroweak
vacuum minimum is less energetically favorable, thus this minimum
decays to the Higgs vacuum, as it was shown in
\cite{Oikonomou:2023bah}. The interesting part though is the phase
transition itself, since such a physical picture destabilizes the
whole dark matter of the Universe even instantaneously. Notice
that this behavior occurs well beyond the neutrino decoupling from
the electroweak sector, and until the temperature of the Universe
is of the order of the neutrino mass. Also there is high
possibility that this first order phase transition may give rise
to gravitational radiation, which is rather unreachable to us via
the interferometric gravitational wave experiments. Perhaps though
it may have some observable effects on the CMB, but this needs
further investigation, which we will not consider in this article
though.

\section{A Qualitative Discussion on the Astrophysical and Cosmological Effects of a Late-time Phase Transition}

In this section we shall qualitatively discuss the astrophysical
and cosmological implications of a late-time first order phase
transition in the Universe. As we shall see, these implications
could be important for the post-CMB Universe and may play a
crucial role on the generation of matter structure at high
redshifts.

The early Universe is known to be very smooth and homogeneous, but
it should have a mechanism that has generated small fluctuations
in the total density of matter and radiation, which eventually
grew into the large scale cosmological structure observed at
present time. The $\Lambda$CDM model, is currently a benchmark
model which predicts that the overall growth of the large scale
structure should unavoidably be influenced by the properties of
cold dark matter, which is currently thought to be the dominant
form of matter in the Universe. In principle, the observations of
the large scale structure of our Universe at high redshifts, which
correspond to an age at which the Universe was quite younger, can
certainly provide important information about the properties of
dark matter at earlier times of our Universe. But these
observations are rather challenging since they require precise
measurements of distant and rather currently faint objects. Recent
observations, for example the ones of the Dark Energy Survey and
the Hyper Suprime-Cam survey \cite{DES:2018ufa}, also provide
constraints on the high redshift large scale structure. Overall,
these observations indicate that the $\Lambda$CDM model is
compatible with the data, but some discrepancies also exist that
must explained. Specifically, the most persisting problem is the
$\sigma_8$ issue having to do with the clustering at high
redshifts with the large scale structure seeming more clustered
than the predictions of the $\Lambda$CDM indicate. This indicates
that the large scale the growth of structure which was formed
after the CMB photons decoupled, was more efficient than the
$\Lambda$CDM model predicts, thus unknown processes, like
late-time phase transitions, may have taken place at the
post-recombination era and these phase transitions may explain
these discrepancies.

Furthermore, regarding the clustering issue, recent observations
coming from the Lyman-alpha forest, which basically is a series of
absorption lines observed in the spectrum of quasars which is
caused by neutral hydrogen gas in the intergalactic medium,
indicate that the clustering properties of these absorption lines
can directly be used to probe the large scale structure of the
Universe. Recent observations of the high redshift Lyman-alpha
forest, indicate that the clustering of matter at high redshifts
is stronger than the predictions of the $\Lambda$CDM model. Also
the clustering of the temperature fluctuations of the CMB  also
indicates a larger clustering of matter, compared to the
predictions of the $\Lambda$CDM model. The creation of large scale
cosmological matter structure is very important, and we shall
further discuss it, in the perspective of late-time first order
phase transitions, which is exactly the type of phase transitions
we presented in this paper. The observational data provided
evidence for high redshift supermassive black holes
\cite{SDSS:2001emm,SDSS:2003iyw,Willott:2003xf,Mortlock:2011va},
and these redshifts correspond to an epoch in which the Universe
was a billion years old. The explanation of the existence of such
massive objects so far back in time is a challenge for the
$\Lambda$CDM model accompanied with a standard inflationary
scenario with Gaussian fluctuations. There exist works in the
literature which discuss such challenges in a concrete way
\cite{Haiman:2000ky,Yoo:2004ze,Shapiro:2004ud,Khlopov:2004sc,Volonteri:2005fj,Volonteri:2006ma,King:2006uu,Li:2006ti,Kawakatu:2009wt,Sijacki:2009mn}.
Late-time phase transitions in the form of first order phase
transitions can generate density fluctuations at relative high
redshifts and thus can explain high redshift structure formation,
without affecting significantly the large-angle CMB temperature
anisotropies. Indeed, detailed studies already exist on this,
which point out that late-time first order phase transitions can
produce non-linear fluctuations that can generate massive
structures of the order $M\sim10^6-10^{10}\,M_{\odot}$, at
redshifts $z\sim 10$ for a phase transition occurring at redshifts
$z\sim 50-500$, see for example \cite{Patwardhan:2014iha}, which
covers our scenario proposed in this paper which predicts phase
transitions in the redshift range $z=[350,37]$. These density
fluctuations which occur due to the first order phase transition,
can be the source of anisotropies on the CMB polarization, and can
be spotted in the integrated Sachs-Wolfe (ISW) effect. The CMB
photons pass through these first order phase transition
originating late-time density fluctuations, before these arrive to
the detectors, and thus the overall effect is the sum of all the
ISW contributions from all the distinct fluctuations. Also we need
to mention the possibility of late-time stochastic gravitational
waves caused by the bubble collisions of the first order phase
transitions, however such a low frequency gravitational radiation
cannot easily be verified experimentally. Several promising
experiments related with the 21cm hydrogen line can in principle
probe high redshift matter formation, up to a redshift of the
order $z\sim 200$
\cite{Loeb:2003ya,Furlanetto:2006jb,Morales:2009gs,Pober:2013jna,Zaldarriaga:2003du,Morales:2012kf,Pritchard:2011xb}.
These low-frequency radio array observations which are based on
the redshift 21cm radiation, can be used to constrain late-time
phase transition scenarios, matter formation and growth for
redshifts up to $z\sim 200$. Late-time phase transitions might
govern early matter formation, growth and clustering, and these
physical scenarios can result to physically distinct situations
compared to the $\Lambda$CDM model. It is notable, that apart from
the fact that the late-time phase transitions can perhaps explain
to some extent the high redshift matter formation and growth, it
is also possible to provide evidence for the evolution and
distribution of dwarf galaxies, which are at the core of
$\Lambda$CDM's shortcomings \cite{Patwardhan:2014iha}.

Late-time phase transitions are linked intrinsically with infrared
physics. Phase transitions are likely to have occurred in the
late-time Universe, at some point after the decoupling of the
photons which form the cosmic microwave background (CMB)
radiation. The reason is that the temperature anisotropy of the
CMB, namely $\delta T/T$ would be affected if phase transitions
occurred before the decoupling of the photons. The CMB last
scattering surface is very smooth, thus if any phase transitions
occurred prior to the decoupling, remnants of this transitions
would disturb the CMB because the phase transitions lead to
curvature perturbations which would have a direct effect on the
CMB last scattering surface. In principle the density fluctuations
at late times can be the source of structure formation at late
times at the post decoupling of photons era, and at redshifts
belonging in the forbidden region of the $\Lambda$CDM, at
redshifts $z\geq 9$ and to mass scales of the order $M\sim
10^{18}M_{\odot}$ \cite{Hill:1988vm}. It is also possible that
soft topological defects may be have been generated during the
late-time phase transitions, which would have cause minimal
effects on the CMB temperature anisotropy, thus the late-time
phase transitions may be an alternative to the inflationary
scenario, as a model that can explain the origin of structure in
the Universe. Thus late-time phase transitions may generate large
scale structure without conflict with the inflationary scenario.
Indeed, such curvature fluctuations generated by late-time phase
transitions may have secondary effects on the temperature
anisotropy of the CMB, which will be generated by the propagation
of the CMB photons through the late-time density fluctuations,
causing a redshifting/blueshifting in the temperature anisotropy.
One should also bare in mind that the late-time phase transitions
may explain ionization and star formation at large redshifts
$z\geq 30$, which could generate ionization. An interesting
feature of late-time phase transitions is that they lead to large
peculiar velocities at the time of the phase transition, due to
large variations in density \cite{Hill:1988vm}. In addition, the
cluster-cluster correlation may be explained by the fractal
character generated from the late-time phase transitions
\cite{Hill:1988vm}.

It is finally worth to discuss the possible implications of a
late-time phase transition on the $H_0$-tension. It is rather
tempting to question the direct effects of the late-time phase
transitions via the axion-neutrino couplings on the $H_0$-tension.
In the context of the present paper, there exists the prediction
of a fifth force among the neutrinos mediated by the axions. It is
known \cite{Kreisch:2019yzn} that strongly interacting 2-to-2
neutrino scatterings with an additional contribution to $\Delta
N_{eff}$ is able to resolve the $H_0$-tension, see also
\cite{Park:2019ibn}. Thus it would be interesting to examine such
a perspective. Also, quite recently the effect the coupling  of
dark matter to neutrinos has been considered \cite{Brax:2023tvn},
in view of the Planck data, and the data hint towards a
non-trivial interaction at 1$\sigma$. See also
\cite{Gerbino:2016sgw} for the impact of neutrinos on inflationary
parameters. An additional motivation to investigate the
implications of a late-time phase transition on the $H_0$-tension
is the fact that such an abrupt physics change at late-times is
known to affect the Cepheid variables
\cite{Perivolaropoulos:2021jda,Perivolaropoulos:2021bds,Perivolaropoulos:2022vql},
see also Refs. \cite{Odintsov:2022eqm,Odintsov:2022umu} for
similar scenarios. It is thus tempting to investigate the effects
of late-time phase transitions on the $H_0$-tension from the
abrupt physics change perspective. Another interesting
perspective, potentially related with the $H_0$-tension, is how a
late-time phase transition may change the expansion rate at the
era of the phase transition, and thus change directly the
calibration of observations related with the era of the phase
transition. Finally, it is important to note that fifth forces,
like the ones predicted in one of the models we presented, are
also connected to the resolution of the $H_0$-tension
\cite{Desmond:2020wep,Sakstein:2019qgn,Desmond:2019ygn} and also
neutrinos are also connected to the $H_0$-tension
\cite{Kreisch:2019yzn,Venzor:2023aka}. For a novel approach on
probing fifth forces, see also \cite{Tsai:2021irw} and also
similar approaches for axion as mediators of the fifth force
between neutrons, see \cite{Capolupo:2021dnl,Capolupo:2023vqb}.

These questions should be answered in a focused and concrete way,
but are out of the scopes of the present article. In the present
section we only sketched on the possible late-time astrophysical
effects of a late-time first order phase transition, without
getting into actual details. Instead of this rather qualitative
approach, detailed hydrodynamical calculations must be performed
in order to see in a quantitative way the actual astrophysical
implications of a late-time phase transitions on all the phenomena
described qualitatively above. This task stretches far beyond the
purposes of this article.

\section{Conclusions}

In this work we considered the possibility of a thermal late-time
phase transition caused by some non-trivial axion-neutrino
interaction which existed primordially. The axion-neutrino system
is supposed to be coupled and the neutrino has a primordial chiral
symmetry which is either broken spontaneously or explicitly. The
two cases of broken chiral symmetry correspond to two distinct
models of axion-neutrinos and we considered the phenomenological
implications of both models. In the case of explicit chiral
symmetry breaking, the axion receives its mass due to one-loop
neutrino corrections and this model is compatible with the strong
naturalness argument. In both cases, the axion is considered to be
the misalignment axion from the primordial era up to the point at
which the first order phase transition might take place, and below
$T\sim m_{\nu}$, the neutrino decouples from the
coherent/classical thermal interaction of neutrino and axion
classical ensembles. Thus the axion acts a cold dark matter when
no phase transition takes place, and it oscillates around the
minimum of its cosine potential. As we demonstrated, there is no
phase transition in the axion-neutrino system, since its effective
potential is not destabilized from the origin due to thermal
corrections, contrary to the existing literature. However, if the
axion interacts with the Higgs particle via higher order
non-renormalizable operators, the axion-neutrino effective
potential is destabilized and a second energetically more
favorable minimum is developed. Thus a first order phase
transition occurs caused by the axion-neutrino interaction, in
which as it seems, the neutrino plays an important role. Thus in
this scenario, the axion acted as cold dark matter from the
primordial era of the Universe, down to the point that the
temperature of the Universe is $T\sim 1\,$MeV, at which point, the
axion is destabilized and a first order phase transition occurs
somewhere in the temperature range $T\sim 0.1-0.01\,$eV, and the
axion vacuum penetrates to the more energetically favorable
vacuum. However, the Higgs vacuum is more energetically favorable
than the axion, and thus the axion minimum decays instantly to the
Higgs vacuum. The late-time phase transition occurs at the
redshift range $z\sim 385-37$, thus it is a post-CMB phase
transition, with no effect on the temperature anisotropy of the
CMB. However, such late-time phase transitions affect the density
fluctuations, which can be sources for late-time structure
formation. Thus high redshift large scale structure and clustering
beyond $z\geq 9$ may be explained by such late-time phase
transitions. It is challenging for astronomers to find hints of
large scale structure at high redshifts and the James Webb Space
Telescope might help towards this perspective. We explored in
detail, in a qualitative way, the phenomenological implications of
late-time first order phase transitions in the axion-neutrino
sector both at the astrophysical and cosmological level. What now
remains is to quantitatively address the issues discussed here
qualitatively, at both cosmological and astrophysical level. This
paper's aim was to point out that the physics of the far infrared
might have more to offer than meets the eye.

\section*{Acknowledgments}

This research has been is funded by the Committee of Science of
the Ministry of Education and Science of the Republic of
Kazakhstan (Grant No. AP19674478).

\end{document}